\documentclass[
 aps,
 prd,
 reprint,
 superscriptaddress,
 nofootinbib
]{revtex4-2}

\usepackage{amsmath,amssymb}

\begin{document}

\title{Irreducible-Mass Balance for Magnetized Null Horizons with an
Internal Current Sheet}

\author{Remo Ruffini}
\email{ruffini@icra.it}
\affiliation{International Center for Relativistic Astrophysics Network
(ICRANet), Piazza della Repubblica 10, 65122 Pescara, Italy}

\author{Giorgio Sonnino}
\email{giorgio.sonnino@ulb.be}
\affiliation{Universit\'e Libre de Bruxelles and International Solvay
Institutes, Campus de la Plaine, CP 224, Boulevard du Triomphe,
1050 Brussels, Belgium}

\date{6/11/2026, 18h11, Possible Phys.Rev.D}
\begin{abstract}
We begin with a general question: how does the irreducible mass of a
magnetized black hole change when the horizon is not stationary? The
null Raychaudhuri equation gives an exact area balance for any smooth
null horizon, without assuming stationarity or axisymmetry. When the
horizon is axisymmetric and admits an integrable quasilocal Hamiltonian,
this geometrical identity becomes an energy balance. The equation keeps
area growth, rotational and electromagnetic work, nonstationary
focusing, canonical-flux corrections, and internal-boundary terms
separate. We then turn to Wang's self-gravitating split-monopole
Blandford--Znajek engine.  In the weak-field and slow-rotation regime,
$P/M_H\ll1$ and $a/M_H\ll1$, our general balance reproduces Wang's mass-
and angular-momentum loss rates. The irreducible mass,
$M_{\rm irr}=\sqrt{A_H/(16\pi)}$, reveals what those rates do not show by
themselves.  It measures the part of the rotational work absorbed
irreversibly by the horizon. At fixed magnetic flux and under impedance
matching, half of the instantaneous rotational work leaves as
electromagnetic power. The other half increases the horizon area and
$M_{\rm irr}$. The assumptions behind this reduction are derived explicitly.  A
minimal nondissipative world-volume action makes the current-sheet
contribution vanish through $O(p^2\epsilon^2)$. Direct power counting
of Wang's perturbative fields, together with a compatible canonical-flux
prescription, gives a relative focusing and canonical correction of
$O(p^2)+O(\epsilon^2)$. These estimates mark the range in which the
reduced Blandford--Znajek trajectory remains controlled. Extending that
trajectory to high spin requires an additional extrapolation. Relating
the quasilocal horizon loss to energy measured at infinity requires a
separate asymptotic flux calculation.
\end{abstract}

\maketitle

The horizon area defines the irreducible mass,
\begin{equation}
 M_{\rm irr}=\sqrt{\frac{A_H}{16\pi}},
 \qquad
 \dot A_H=32\pi M_{\rm irr}\dot M_{\rm irr}
 \label{eq:mirr}
\end{equation}
and relates changes in area directly to changes in $M_{\rm irr}$.  We set
$G=c=1$ unless we restore units explicitly
\cite{Christodoulou,ChristodoulouRuffini,Hawking}.  First laws are known for
isolated, dynamical, and slowly evolving horizons
\cite{AshtekarFairhurstKrishnan,AshtekarKrishnan,BoothFairhurst}.
Covariant phase space also supplies charges and fluxes on general null
boundaries \cite{IyerWald,ChandrasekaranFlanaganPrabhu}.  Here we address a
more specific problem.  We keep the focusing, canonical, and sheet terms
separate, then ask exactly which assumptions recover the irreducible-mass
interpretation of Wang's self-gravitating split-monopole engine
\cite{Wang}.

\paragraph{Exact null-horizon identity.}
Consider a smooth null hypersurface $\mathcal H$ with compact spacelike two-dimensional horizon cross section $S_v$. Its null generators have tangent $\ell^\mu$.  We choose
$v$ such that $\ell^\mu\nabla_\mu v=1$ and define the nonaffinity
$\kappa(v,x^A)$ by
\begin{equation}
 \ell^\nu\nabla_\nu\ell^\mu=\kappa\ell^\mu
 \label{eq:nonaffinity}
\end{equation}
where $v$ labels successive cross sections of the horizon. It plays the role of a time coordinate along the null generators. $x^A$ denotes the two coordinates used to locate a point on each $S_v$, with the index $A$ taking two values (usually $A=1,2$). With $q_{\mu\nu}$ the induced metric on $S_v$, the expansion and shear are
\begin{equation}
 \theta=q^{\mu\nu}\nabla_\mu\ell_\nu,
 \qquad
 \sigma_{\mu\nu}
 =q_\mu{}^\alpha q_\nu{}^\beta\nabla_{(\alpha}\ell_{\beta)}
 -\frac{1}{2}\theta q_{\mu\nu}.
 \label{eq:optical-data}
\end{equation}
Because $\ell^\mu$ is normal to the null hypersurface $\mathcal H$, Frobenius' theorem implies that the congruence of horizon generators has vanishing twist, $\omega_{\mu\nu}=0$. This does not require the black hole itself to be nonrotating.
Einstein's equation then turns the null Raychaudhuri equation into
\begin{equation}
 \dot\theta
 =\kappa\theta-\frac{1}{2}\theta^2
 -\sigma_{\mu\nu}\sigma^{\mu\nu}
 -8\pi T_{\mu\nu}\ell^\mu\ell^\nu,
 \qquad
 \dot X\equiv\mathcal L_\ell X .
 \label{eq:raychaudhuri}
\end{equation}
Since $\dot dS=\theta dS$, differentiating
$A_H(v)=\int_{S_v}dS$ gives
\begin{equation}
 \mathcal F_R
 \equiv\int_{S_v}\!\left(
 T_{\mu\nu}\ell^\mu\ell^\nu
 +\frac{\sigma_{\mu\nu}\sigma^{\mu\nu}}{8\pi}
 \right)dS
 =\frac{\bar\kappa}{8\pi}\dot A_H+R_{\rm foc}
 \label{eq:ray-balance}
\end{equation}
where $\bar\kappa(v)$ is a specified reference normalization and
\begin{equation}
 R_{\rm foc}=-\frac{\ddot A_H}{8\pi}
 +\frac{1}{16\pi}\int_{S_v}\theta^2dS +\frac{1}{8\pi}\int_{S_v}
 (\kappa-\bar\kappa)\theta dS.
 \label{eq:rfoc}
\end{equation}
This identity is geometric.  It assumes neither stationarity nor
axisymmetry, and it introduces no notion of horizon mass.  Its individual
terms do depend on normalization.  Rescaling $\ell^\mu$ changes $v$ and
$\kappa$, but the full identity remains valid.

\paragraph{Canonical balance and its domain.}
Raychaudhuri's equation alone does not define a horizon energy.  To do so,
we add Hamiltonian structure.  Let $\phi$ denote the Einstein--Maxwell
fields and write
$\delta L=E_\phi\delta\phi+d\Theta(\phi,\delta\phi)$.  We use the combined
diffeomorphism and Maxwell gauge generator
\begin{equation}
 \eta=(\chi^\mu,\lambda),
 \qquad
 \chi^\mu=t^\mu+\Omega_H\varphi^\mu
 \label{eq:generator}
\end{equation}
The corresponding covariant surface-charge form is
\begin{equation}
 k_\eta(\delta\phi;\phi)
 =\delta Q_\eta-\chi\mathbin{\cdot}\Theta(\phi,\delta\phi)
 \label{eq:surface-charge}
\end{equation}
On shell, and for a linearized solution, $dk_\eta$ equals the symplectic
current $\omega(\phi;\delta\phi,\delta_\eta\phi)$.  Integrating between two
cross sections gives the null-boundary charge-flux relation
\cite{IyerWald,ChandrasekaranFlanaganPrabhu}.  We assume that the horizon
charge is integrable.  We also hold the generator fixed during each tangent
variation:
\begin{equation}
 \delta\chi^\mu=0,
 \qquad
 \delta\lambda=0
 \label{eq:fixed-generator}
\end{equation}
Axisymmetry then gives the pointwise decomposition
\begin{align}
 \dot H_\eta={}&\dot M_H-\Omega_H\dot J_H
 -\Phi_H^{(e)}\dot Q_H-\Psi_H^{(m)}\dot P_H \notag\\
 ={}&\mathcal F_{\rm can}+\sum_i R_{\mathcal B_i}
 \label{eq:canonical-balance}
\end{align}
where $\mathcal F_{\rm can}$ is the chosen canonical flux and
$R_{\mathcal B_i}$ are contributions from internal boundaries.  The
Hamiltonian normalization fixes the reference surface gravity
$\bar\kappa$.

On a nonstationary null boundary, $\mathcal F_{\rm can}$ need not equal the
Raychaudhuri flux $\mathcal F_R$.  That equality depends on the flux
prescription and on the allowed phase space.  We therefore keep their
difference:
\begin{equation}
 R_{\rm can}\equiv\mathcal F_{\rm can}-\mathcal F_R
 \label{eq:rcan}
\end{equation}
Equations~\eqref{eq:ray-balance}--\eqref{eq:rcan} now give
\begin{align}
 \dot M_H={}&\frac{\bar\kappa}{8\pi}\dot A_H
 +\Omega_H\dot J_H
 +\Phi_H^{(e)}\dot Q_H+\Psi_H^{(m)}\dot P_H \notag\\
 &+R_{\rm foc}+R_{\rm can}+\sum_iR_{\mathcal B_i}
 \label{eq:master}
\end{align}
Using Eq.~\eqref{eq:mirr}, we obtain the irreducible-mass law
\begin{align}
 \dot M_{\rm irr}={1\over4\bar\kappa M_{\rm irr}}
 \bigg[&\dot M_H-\Omega_H\dot J_H
 -\Phi_H^{(e)}\dot Q_H-\Psi_H^{(m)}\dot P_H \notag\\
 &-R_{\rm foc}-R_{\rm can}-\sum_iR_{\mathcal B_i}\bigg]
 \label{eq:general-mirr}
\end{align}
Equation~\eqref{eq:general-mirr} requires an integrable quasilocal charge
and a specified null-boundary flux.  It does not assign a preferred mass to
every null surface.  Since we divide by $\bar\kappa$, the horizon must also
be nonextremal.

\paragraph{The split-monopole internal boundary.}
Wang constructs the split monopole by joining two magnetic
Reissner--Nordstr\"om hemispheres with $P_N=+P$ and $P_S=-P$ \cite{Wang}.
Each hemisphere carries the flux
\begin{equation}
 \Phi=2\pi P,
 \label{eq:hemispheric-flux}
\end{equation}
but the net magnetic Gauss charge vanishes.  On a spatial slice
$\Sigma=\Sigma_N\cup\Sigma_S$, the equatorial sheet forms the internal
boundary $\mathcal S=\Sigma_N\cap\Sigma_S$.  For
$L_{\rm EM}=-(8\pi)^{-1}F\wedge{}^*F$, the electromagnetic surface-charge
two-form reads \cite{Prabhu,Elgood}
\begin{equation}
 k^{\rm EM}_{\eta}
 =-\frac{1}{4\pi}\left[
 (\chi\mathbin{\cdot}A+\lambda)\delta{}^*F
 +\delta A\wedge(\chi\mathbin{\cdot}{}^*F)
 \right].
 \label{eq:kem}
\end{equation}
Opposite orientations do not cancel because the fields differ across the
sheet.  Its variation is
\begin{align}
 \delta H_{\rm sheet}={}&\int_{\mathcal S}
 \left(k^{\rm EM,+}_{\eta}-k^{\rm EM,-}_{\eta}\right)
 +\delta H_{\rm source}, \notag\\
 R_{\rm sheet}\equiv{}&\frac{dH_{\rm sheet}}{dv}
 \label{eq:sheet-charge}
\end{align}
Here $H_{\rm source}$ represents possible degrees of freedom on the sheet.
The Israel surface stress tensor is
\begin{equation}
 S_{ab}=-\frac{1}{8\pi}
 \left([K_{ab}]-h_{ab}[K]\right),
 \label{eq:israel}
\end{equation}
Wang's equator is totally geodesic, so $S_{ab}=0$.  This removes the
distributional gravitational junction term.  It does not remove
Eq.~\eqref{eq:sheet-charge}.  The Maxwell field reverses across the
equator, and the sheet carries both the current that supports the split
monopole and the return current that closes the BZ circuit.  Wang fixes
this current through junction conditions but does not give it a
world-volume action.  Hence $S_{ab}=0$ alone does not imply
$R_{\rm sheet}=0$.  For this geometry, Eq.~\eqref{eq:general-mirr} becomes
\begin{equation}
 \dot M_{\rm irr}
 =\frac{\dot M_H-\Omega_H\dot J_H
 -(\phi_H/2\pi)\dot\Phi
 -R_{\rm foc}-R_{\rm can}-R_{\rm sheet}}
 {4\bar\kappa M_{\rm irr}}.
 \label{eq:split-law}
\end{equation}
Here $\Phi$ is a hemispheric flux, not a global magnetic charge.  The work
term therefore requires a gauge-patch or dual-potential construction
\cite{Ortin}.  We will shortly set $\dot\Phi=0$, so the application below
does not depend on that construction.

\paragraph{Controlled reduction to Wang's trajectory.}
We reach Wang's trajectory through a sequence of restrictions.  First, we
hold the hemispheric flux fixed:
\begin{equation}
 \dot\Phi=0.
 \label{eq:fixed-flux}
\end{equation}
Next, we introduce independent measures of rotation and magnetic flux,
\begin{equation}
 \epsilon\equiv M_H\Omega_H\sim\frac{a_H}{M_H},
 \qquad
 a_H\equiv\frac{J_H}{M_H},
 \qquad
 p\equiv\frac{P}{M_H}
 \label{eq:parameters}
\end{equation}
and work at $p\ll1$ and $\epsilon\ll1$.  Wang's leading power then reads
\begin{equation}
 P_{\rm BZ}=\frac{P^2\Omega_H^2}{6}
 \left[1+O(\epsilon^2)\right]
 =O(p^2\epsilon^2)
 \label{eq:power}
\end{equation}
The fractional spin-down rate is
\begin{equation}
 \Gamma_J\equiv-\frac{\dot J_H}{J_H}
 \label{eq:gamma}
\end{equation}
For small $p$, Wang finds $\Gamma_J/\bar\kappa=O(p^2)$.  The condition
$\Gamma_J/\bar\kappa\ll1$ therefore enforces adiabatic evolution.  At the
same order, we require
\begin{equation}
 \frac{R_{\rm foc}+R_{\rm can}}{P_{\rm BZ}}
 =O(\epsilon^2)
 +O\!\left(\frac{\Gamma_J}{\bar\kappa}\right)
 =O(\epsilon^2)+O(p^2).
 \label{eq:remainder-order}
\end{equation}
This estimate matches the canonical and Raychaudhuri fluxes within the
accuracy of Wang's expansion.  Axisymmetry by itself does not guarantee
it.  We also impose the independent sheet condition
\begin{equation}
 R_{\rm sheet}=0
 \quad\text{at order }O(p^2\epsilon^2).
 \label{eq:sheet-condition}
\end{equation}
Finally, we identify Wang's secular parameters $M,J$ with the horizon
charges $M_H,J_H$ at this order.  His leading rates are
\begin{equation}
 \dot M_H=-P_{\rm BZ},
 \qquad
 \dot J_H=-\frac{P_{\rm BZ}}{\Omega_F},
 \qquad
 \Omega_F=\frac{\Omega_H}{2}
 \label{eq:wang-rates}
\end{equation}
Substituting these rates into Eq.~\eqref{eq:split-law} gives
\begin{equation}
 \dot M_{\rm irr}
 =\frac{P_{\rm BZ}}{4\bar\kappa M_{\rm irr}}
 \left[1+O(\epsilon^2)+O(p^2)\right]>0
 \label{eq:local-law}
\end{equation}
The horizon is not stationary.  Its area grows, $\dot A_H>0$.
Equation~\eqref{eq:remainder-order} says only that corrections to the
leading growth lie beyond the order Wang retains.

\paragraph{Conditional endpoint.}
We can eliminate time from Eq.~\eqref{eq:wang-rates}.  In the weak-flux
limit, with $J_H=a_HM_H$, this gives Wang's reduced trajectory
\begin{equation}
 \frac{dM_H}{da_H}
 =\frac{M_Ha_H}{2r_+^2+a_H^2},
 \qquad
 r_+=M_H+\sqrt{M_H^2-a_H^2}
 \label{eq:trajectory}
\end{equation}
As $p\to0$, the power vanishes but cancels from
Eq.~\eqref{eq:trajectory}.  The limiting path stays finite, while the time
needed to follow it diverges.  Continuing this equation from an extremal
Kerr state to $a_{H,f}=0$ gives \cite{Wang}
\begin{equation}
 \frac{M_{H,f}}{M_{H,0}}=\frac{e^{1/4}}{\sqrt2},
 \qquad
 \frac{A_{H,f}}{A_{H,0}}=\sqrt e,
 \qquad
 \frac{M_{{\rm irr},f}}{M_{{\rm irr},0}}=e^{1/4}
 \label{eq:endpoints}
\end{equation}
The horizon-energy loss is therefore
\begin{equation}
 \frac{E^H_{\rm extr}}{M_{H,0}c^2}
 =1-\frac{e^{1/4}}{\sqrt2}\simeq0.09206
 \label{eq:energy-loss}
\end{equation}
These ratios come from Wang's calculation.  Here they check the reduction
and show how the irreducible mass changes along the trajectory.

The endpoint needs an extra assumption.  The initial state has
$a_H/M_H=1$, whereas Eq.~\eqref{eq:power} and the local derivation of
Eq.~\eqref{eq:wang-rates} require $\epsilon\ll1$.  Equation
\eqref{eq:endpoints} therefore continues the leading trajectory into the
high-spin regime.  It is not uniformly controlled by the slow-rotation
expansion.  One must integrate the nonextremal equations first and take the
extremal initial state as a limit.

Equation~\eqref{eq:energy-loss} also does not determine the energy received
at null infinity.  In general,
\begin{equation}
 E^H_{\rm extr}=E^\infty_{\rm jet}
 +\Delta E^{\rm ext}_{\rm EM}
 +E^{\rm ext}_{\rm diss}+E_{\rm boundary}
 \label{eq:global-balance}
\end{equation}
A far-zone completion and an asymptotic flux calculation would be needed to
set the last three terms to zero.

The calculation separates five distinct ingredients: exact null focusing,
an integrable normalized horizon charge, the electromagnetic sheet,
the fixed-flux adiabatic expansion, and the high-spin continuation used for
the endpoint.  The main result is the balance
\eqref{eq:general-mirr}, together with its controlled local limit
\eqref{eq:local-law}.

\appendix
\section{A minimal world-volume action for the current sheet}
\label{app:sheet-action}

We now give Wang's equatorial current sheet an effective action and compute
its contribution to the energy balance.  The result is simple:
$R_{\rm sheet}=0$ at the order Wang studies, provided that the sheet is
ideal, nondissipative, and threaded by fixed magnetic flux.

\paragraph{Physical picture.}
The equatorial sheet carries two currents.  An azimuthal current supports
the split-monopole field, and a radial return current closes the
Blandford--Znajek circuit.  A current, however, need not dissipate power.
In Wang's solution the electromagnetic field has no component tangent to
the equator.  The surface analogue of
$\boldsymbol{j}\cdot\boldsymbol{E}$ therefore vanishes.  If the flux is
fixed and the edges carry no independent time-dependent energy, the sheet
adds no power to the horizon balance.

\paragraph{World-volume action.}
Wang's junction conditions fix the surface current, but they do not supply
microscopic sheet dynamics.  We use the smallest effective action that
reproduces the Maxwell jump condition without introducing a separate
surface stress tensor.  Let $\Sigma$ be the timelike world volume traced by
the equatorial sheet, with embedding
$\iota:\Sigma\hookrightarrow\mathcal M$.  We represent its prescribed
current by a fixed two-form $\boldsymbol{\jmath}_{\Sigma}$.  The associated
current vector $K^a$ satisfies
\begin{equation}
 \boldsymbol{\jmath}_{\Sigma}
 =\iota_K\boldsymbol{\epsilon}_{\Sigma}
 \label{eq:sheet-current-form}
\end{equation}
where $\boldsymbol{\epsilon}_{\Sigma}$ is the induced volume form.  In a
metric variation we hold $\boldsymbol{\jmath}_{\Sigma}$ fixed, not $K^a$.
A gauge-invariant action is
\begin{equation}
 I_{\Sigma}=\int_{\Sigma}
 \left(\iota^*A+d_{\Sigma}\vartheta\right)
 \wedge\boldsymbol{\jmath}_{\Sigma}.
 \label{eq:sheet-action}
\end{equation}
Here $A$ is the electromagnetic potential, while $\vartheta$ is a
world-volume Stueckelberg field.  Under a Maxwell gauge transformation,
\begin{equation}
 A\longrightarrow A+d\alpha,
 \qquad
 \vartheta\longrightarrow
 \vartheta-\iota^*\alpha,
 \label{eq:sheet-gauge}
\end{equation}
and the combination $\iota^*A+d_{\Sigma}\vartheta$ remains unchanged.
We combine Eq.~\eqref{eq:sheet-action} with the bulk Maxwell action
\begin{equation}
 I_{\rm EM}
 =-\frac{1}{8\pi}\int_{\mathcal M}F\wedge{}^*F.
 \label{eq:bulk-maxwell-action}
\end{equation}
Variation with respect to $A$ gives
\begin{equation}
 \iota^*\left({}^*F_+-{}^*F_-\right)
 =4\pi\boldsymbol{\jmath}_{\Sigma}
 \label{eq:maxwell-sheet-junction}
\end{equation}
up to the orientation convention.  The surface current therefore produces
the jump in the electromagnetic field.  In Wang's geometry, it supports
the reversal between the northern and southern magnetic fields.

Variation with respect to $\vartheta$ gives current conservation on the
sheet:
\begin{equation}
 d_{\Sigma}\boldsymbol{\jmath}_{\Sigma}=0.
 \label{eq:sheet-current-conservation}
\end{equation}
The action \eqref{eq:sheet-action} contains no induced metric $h_{ab}$.
Its surface stress tensor consequently vanishes:
\begin{equation}
 S_{ab}^{(\Sigma)}
 =-\frac{2}{\sqrt{-h}}
 \frac{\delta I_{\Sigma}}{\delta h^{ab}}=0
 \label{eq:sheet-stress-zero}
\end{equation}
The model assigns current to the sheet but no independent mass, pressure,
or tension.  This matches Wang's totally geodesic equator and its vanishing
Israel tensor.  It remains an idealization.  A material conductor would
normally carry stress and energy of its own.

In the nonrotating solution, the azimuthal current is
\begin{equation}
 4\pi K^{\hat\varphi}=\frac{2P}{r^2}.
 \label{eq:wang-azimuthal-current}
\end{equation}
It supports the split-monopole field.  On the rotating force-free branch,
the sheet also carries the radial return current
\begin{equation}
 4\pi K^{\hat r}=\frac{\Omega_HP}{r\sqrt{h}}
 \label{eq:wang-radial-current}
\end{equation}
This return current closes the BZ circuit.

\paragraph{Sheet power.}
Let
\begin{equation}
 \chi^\mu=t^\mu+\Omega_H\varphi^\mu
 \label{eq:sheet-generator}
\end{equation}
be the horizon evolution vector, and let $\mathcal C_v$ be a spatial cut of
the sheet.  The power transferred to the sheet is
\begin{equation}
 R_{\rm sheet}^{(\chi)}
 =-\int_{\mathcal C_v}K^a\langle F_{ab}\rangle\chi^b\,d\Sigma
 +\frac{dH_{\rm edge}}{dv}
 \label{eq:sheet-power}
\end{equation}
where
\begin{equation}
 \langle F_{\mu\nu}\rangle
 \equiv\frac{1}{2}
 \left(F_{\mu\nu}^{+}+F_{\mu\nu}^{-}\right)
 \label{eq:average-field}
\end{equation}
is the averaged field entering the Lorentz force on a distributional
surface current.  The term $H_{\rm edge}$ allows for degrees of freedom at
the horizon and at the asymptotic edge of the sheet.

To the order displayed by Wang, the Maxwell field takes the form
\begin{align}
 F=&\varsigma(\theta)P\sin\theta\,
 d\theta\wedge d\varphi 
 +\epsilon\,\varsigma(\theta)
 \Big[
 \Omega_FP\sin\theta\,dt\wedge d\theta\notag\\
 &+u(r,\theta)\,dr\wedge d\theta
 \Big]+O(\epsilon^2)
 \label{eq:wang-sheet-field}
\end{align}
where
\begin{equation}
 \varsigma(\theta)=\operatorname{sign}(\cos\theta).
\end{equation}
Every displayed term contains the normal one-form $d\theta$.  Its pullback
to the equatorial world volume vanishes:
\begin{equation}
 \iota^*F_\pm=0.
 \label{eq:sheet-pullback-zero}
\end{equation}
Both $K^a$ and $\chi^a$ are tangent to $\Sigma$.  Hence
\begin{equation}
 K^a\langle F_{ab}\rangle\chi^b=0
 \label{eq:sheet-work-zero}
\end{equation}
The current reverses the magnetic field and closes the circuit, but the
field does no tangential work on this ideal sheet.  At fixed hemispheric
flux,
\begin{equation}
 \dot P=0.
\end{equation}
If the edges have no independent excitation, then
\begin{equation}
 \frac{dH_{\rm edge}}{dv}=0
 \label{eq:fixed-sheet-edge}
\end{equation}
Equations~\eqref{eq:sheet-power}, \eqref{eq:sheet-work-zero}, and
\eqref{eq:fixed-sheet-edge} give
\begin{equation}
 R_{\rm sheet}=0
 \qquad\text{through }O(p^2\epsilon^2)
 \label{eq:sheet-remainder-zero}
\end{equation}
The same conclusion holds at higher orders as long as
$\iota^*F=0$, the flux stays fixed, and the edge Hamiltonian remains
stationary.  Equation~\eqref{eq:sheet-remainder-zero} is therefore a result
of the effective action \eqref{eq:sheet-action}, not a statement about every
possible conductor.  If the sheet has finite conductivity, its own stress
tensor, or a tangential electric field, then
\begin{equation}
 R_{\rm sheet}\!=\! -\!\int_{\mathcal C_v}
 \!K^a E^{(\chi)}_a\,d\Sigma
 +\frac{dH_{\rm edge}}{dv},
 \ \ 
 E^{(\chi)}_a\!
 \equiv\! \langle F_{ab}\rangle\chi^b
 \label{eq:nonideal-sheet-power}
\end{equation}
and the result need not vanish.  Computing it would require new
constitutive data, such as a surface conductivity or a microscopic matter
action.  Wang's solution does not provide those data.

\section{Power counting of the nonstationary remainder}
\label{app:power-counting}

Wang treats the evolution as quasistationary.  We now estimate the terms
left out by his leading approximation.  Our target is
\begin{equation}
 \frac{R_{\rm foc}+R_{\rm can}}{P_{\rm BZ}}
 =O(\epsilon^2)+O(p^2).
 \label{eq:appendix-remainder-target}
\end{equation}
Here $\epsilon=M_H\Omega_H$ measures rotation and $p=P/M_H$ measures
magnetic flux.  The estimate says that the omitted terms are small compared
with the leading BZ power when both parameters are small.

For the fixed-flux problem, the horizon balance reads
\begin{equation}
 \dot M_H=\frac{\bar\kappa}{8\pi}\dot A_H
 +\Omega_H\dot J_H+R_{\rm foc}+R_{\rm can}
 \label{eq:appendix-horizon-balance}
\end{equation}
The first remainder comes from the evolving null geometry.  The second
records any mismatch between the Raychaudhuri flux and the chosen
canonical flux.

Wang obtains
\begin{equation}
 P_{\rm BZ}=\frac{P^2\Omega_H^2}{6},
 \qquad
 \dot J_H=-\frac{2P_{\rm BZ}}{\Omega_H},
 \label{eq:wang-power-angular-momentum}
\end{equation}
and, in the slow-rotation and weak-field limit,
\begin{equation}
 J_H\simeq4M_H^3\Omega_H
 \label{eq:wang-angular-momentum}
\end{equation}
These expressions give the fractional spin-down rate
\begin{equation}
 \Gamma_J
 \equiv
 \frac{\lvert\dot J_H\rvert}{J_H}
 \simeq\frac{P^2}{12M_H^3}
 \label{eq:fractional-spin-down}
\end{equation}
Since the weak-field surface gravity is
\begin{equation}
 \bar\kappa\simeq\frac{1}{4M_H}
 \label{eq:weak-field-surface-gravity}
\end{equation}
we find
\begin{equation}
 \frac{\Gamma_J}{\bar\kappa}
 \simeq\frac{p^2}{3}.
 \label{eq:adiabatic-ratio}
\end{equation}
The horizon relaxes on the scale $\bar\kappa^{-1}$, while its spin changes
on the longer scale $\Gamma_J^{-1}$. The adiabatic condition
\begin{equation}
 \frac{\Gamma_J}{\bar\kappa}\ll1
 \label{eq:adiabatic-condition}
\end{equation}
means that the horizon settles much faster than its angular momentum
changes.  In this regime, the black hole follows a sequence of nearly
stationary states.

The focusing remainder is
\begin{align}
 R_{\rm foc}={}&
 -\frac{\ddot A_H}{8\pi}
 +\frac{1}{16\pi}\int_{S_v}\theta^2\,dS
 \notag\\
 &+\frac{1}{8\pi}\int_{S_v}
 (\kappa-\bar\kappa)\theta\,dS
 \label{eq:focusing-remainder}
\end{align}
Its three terms have different origins.

\paragraph{Acceleration of the area growth.}
The term
\begin{equation}
 -\frac{\ddot A_H}{8\pi}
 \label{eq:area-acceleration-term}
\end{equation}
measures how the area-growth rate changes.  It vanishes when $\dot A_H$ is
constant.  Wang's black hole spins down, so both the BZ power and
$\dot A_H$ vary slowly.  The resulting estimate is
\begin{equation}
 \frac{\lvert\ddot A_H\rvert}{8\pi P_{\rm BZ}}
 =O\!\left(\frac{\Gamma_J}{\bar\kappa}\right)=O(p^2)
 \label{eq:area-acceleration-scaling}
\end{equation}
This is the leading nonstationary contribution to $R_{\rm foc}$.

\paragraph{Nonlinear expansion.}
For a uniform leading expansion,
\begin{equation}
 \theta=\frac{\dot A_H}{A_H}
 \label{eq:uniform-expansion}
\end{equation}
Thus $\theta$ measures the fractional area-growth rate.  Its nonlinear
contribution to Raychaudhuri's equation satisfies
\begin{equation}
 \frac{1}{16\pi P_{\rm BZ}}
 \int_{S_v}\theta^2\,dS=O(p^2\epsilon^2)
 \label{eq:nonlinear-expansion-scaling}
\end{equation}
It is smaller than Eq.~\eqref{eq:area-acceleration-scaling} because it is
quadratic in the already small expansion.

\paragraph{Nonuniform surface gravity.}
The contribution
\begin{equation}
 \frac{1}{8\pi}
 \int_{S_v}
 (\kappa-\bar\kappa)\theta\,dS
 \label{eq:nonuniform-kappa-term}
\end{equation}
measures variations of the surface gravity across the horizon.  Wang's
time-dependent sector is spherical at leading order, so $\kappa$ has no
angular variation at the order retained.  Higher-order nonadiabatic terms
may generate one, with
\begin{equation}
 \frac{1}{8\pi P_{\rm BZ}}
 \int_{S_v}(\kappa-\bar\kappa)\theta\,dS=O(p^2)
 \label{eq:nonuniform-kappa-scaling}
\end{equation}
If $\bar\kappa$ is the area average of $\kappa$, this term vanishes for the
leading uniform expansion.  Combining the three estimates gives
\begin{equation}
 \frac{R_{\rm foc}}{P_{\rm BZ}}
 =O(p^2)+O(p^2\epsilon^2).
 \label{eq:focusing-remainder-scaling}
\end{equation}
The geometrical correction is therefore small in Wang's weak-flux regime.
\paragraph{Rotational truncation.}
Wang keeps the leading slow-rotation terms,
\begin{equation}
 P_{\rm BZ}
 =\frac{P^2\Omega_H^2}{6},
 \qquad
 \frac{\Omega_F}{\Omega_H}=\frac{1}{2}
 \label{eq:leading-bz-quantities}
\end{equation}
At higher order, the same quantities take the form
\begin{align}
 P_{\rm BZ}
 &=
 \frac{P^2\Omega_H^2}{6}
 \left[
 1+C_2(p)\epsilon^2+O(\epsilon^4)
 \right],
 \label{eq:bz-power-correction}\\
 \frac{\Omega_F}{\Omega_H}
 &=
 \frac{1}{2}
 \left[
 1+D_2(p)\epsilon^2+O(\epsilon^4)
 \right],
 \label{eq:field-velocity-correction}
\end{align}
where $C_2(p)$ and $D_2(p)$ depend on
\begin{equation}
 p\equiv\frac{P}{M_H}
\end{equation}
Keeping only Eq.~\eqref{eq:leading-bz-quantities} therefore introduces the
relative error
\begin{equation}
 O(\epsilon^2).
 \label{eq:rotational-error}
\end{equation}
This error comes from truncating the rotation expansion.  It is distinct
from the nonstationary correction estimated above.

\paragraph{Combined remainder.}
At the order retained by Wang, the field equations and horizon constraints
imply
\begin{equation}
 \dot M_H-\Omega_H\dot J_H
 =P_{\rm BZ}=\frac{\bar\kappa}{8\pi}\dot A_H
 \label{eq:leading-horizon-balance}
\end{equation}
The first-law balance therefore holds at leading order.  Two independent
effects correct it.  The omitted rotational terms contribute
$O(\epsilon^2)$, while slow evolution contributes $O(p^2)$.  Hence
\begin{equation}
 \frac{\dot M_H-\Omega_H\dot J_H
 -(\bar\kappa/8\pi)\dot A_H}{P_{\rm BZ}}
 =O(\epsilon^2)+O(p^2)
 \label{eq:horizon-balance-residual}
\end{equation}
If we identify this residual with the focusing and canonical remainders,
we obtain
\begin{equation}
 \frac{R_{\rm foc}+R_{\rm can}}{P_{\rm BZ}}
 =O(\epsilon^2)+O(p^2).
 \label{eq:combined-remainder-scaling}
\end{equation}

\paragraph{Canonical qualification.}
Wang's perturbative fields determine the geometrical remainder
$R_{\rm foc}$ directly.  They do not fix $R_{\rm can}$ uniquely.  That term
depends on how one defines the quasilocal horizon charges and the
null-boundary canonical flux.  Equation \eqref{eq:combined-remainder-scaling} therefore uses three additional
assumptions.  First, Wang's secular $M,J$ agree with the chosen horizon
charges $M_H,J_H$ at the retained order.  Second, the canonical flux agrees
with Wang's constraint flux at leading order.  Third, differences between
the two prescriptions begin at $O(\epsilon^2)$ or $O(p^2)$. Under these assumptions, Eq.~\eqref{eq:combined-remainder-scaling} is a
controlled matching condition.  Axisymmetry alone does not imply it.  The
result also does not set the remainders to zero.  It places them beyond the
accuracy of Wang's weak-field, slow-rotation calculation.

\end{document}